\documentclass[aps,prb,twocolumn,showpacs,superscriptaddress]{revtex4}
\usepackage{amssymb,amsmath}
\usepackage{graphicx}
\usepackage{multirow}
\usepackage{hyperref}
\usepackage{appendix}
\usepackage{mathtools}
\usepackage{amsmath}
\usepackage{xcolor}
\usepackage{siunitx}
\graphicspath{{./figs/}}

\begin{document}

\title{
Design rule for the emission linewidth of Eu$^{2+}$-activated phosphors}

\author{Yongchao Jia}
\email[]{yongchao.jia@uclouvain.be}
\affiliation{European Theoretical Spectroscopy Facility, Institute of Condensed Matter and Nanosciences, Universit\'{e} catholique de Louvain, Chemin des \'{e}toiles 8, bte L07.03.01, B-1348 Louvain-la-Neuve, Belgium}
\author{Samuel Ponc\'{e}}
\affiliation{Theory and Simulation of Materials (THEOS), \'Ecole Polytechnique F\'ed\'erale de Lausanne, CH-1015 Lausanne, Switzerland}
\author{Anna Miglio}
\affiliation{European Theoretical Spectroscopy Facility, Institute of Condensed Matter and Nanosciences, Universit\'{e} catholique de Louvain, Chemin des \'{e}toiles 8, bte L07.03.01, B-1348 Louvain-la-Neuve, Belgium}
\author{Masayoshi Mikami}
\affiliation{Functional Materials Design Laboratory, Yokohama R$\&$D Center, 1000,
Kamoshida-cho Aoba-ku, Yokohama, 227-8502, Japan}
\author{Xavier Gonze}
\affiliation{European Theoretical Spectroscopy Facility, Institute of Condensed Matter and Nanosciences, Universit\'{e} catholique de Louvain, Chemin des \'{e}toiles 8, bte L07.03.01, B-1348 Louvain-la-Neuve, Belgium}
\affiliation{Skolkovo Institute of
Science and Technology, Skolkovo Innovation Center, Nobel St. 3, Moscow, 143026, Russia.}

\begin{abstract}
We study from first principles the emission linewidth of Eu$^{2+}$-doped LED phosphors. 
Based on the one-dimensional configuration coordinate model, an analysis of first principles data
obtained for fifteen compounds show that, at working temperature, the linewidth of Eu$^{2+}$ emission band in solids is negligibly affected by quantum effects, 
and can be extracted from the Franck-Condon energy shifts. 
For a fixed Stokes shift, the difference of Franck-Condon energy shifts in the excited and ground states is the key factor for the FWHM determination.
Narrow emission Eu$^{2+}$-doped LED phosphors are expected for the case with large positive value of such difference. 

\end{abstract}




\maketitle

\section{Introduction}
\label{intro}

Due to the coupling of electronic transition with lattice vibrations, the photoluminescence of color centers in solids naturally appears as vibronic spectra, composed by a zero-phonon line and phonon sidebands, whose ratio decreases as the coupling increases\cite{huang1950theory,rebane2012impurity,henderson2006optical}. 
The integrated intensity of these spectrum lines is independent of temperature in the Born adiabatic approximations\cite{born1954dynamical}.
However, as temperature increases, spectral weight is transferred from the zero-phonon line to the phonon sidebands. 
The electron-phonon coupling significantly increases the width of photoluminescence spectra of color centers compared to their intrinsic widths, which depends on electronic transition lifetime\cite{rebane2012impurity}.

The broad character of the photoluminescence spectrum has recently found many commercial applications.
A paradigmatic example is the white light-emitting diode (LED), that relies on the 4f$\leftrightarrow$5d transition of Ce$^{3+}$ and Eu$^{2+}$ ions, converting  high energy photons into targeted color and creating white light\cite{shionoya2018phosphor,jia2012color}. 
The 4f$\leftrightarrow$5d transition is characterized by an intense electron-phonon coupling, since the 5d electronic state of Ce$^{3+}$ and Eu$^{2+}$ ions has a large spatial extension and strongly feels its crystal environment. 
The emission linewidth of LED phosphors is an important factor determining the performance and tunability of optical devices.

Recently, the US Department of Energy has listed the requirements for red-emitting Eu$^{2+}$ ions-doped phosphors for warm white light generation\cite{2019-ssl}. 
At the working temperature of LED devices, the full-width at half maximum (FWHM) of red phosphors should be smaller than 30~nm to achieve better optical quality.
Following this requirement, many efforts have been devoted to this topic, among which the discovery of a promising red phosphor Sr[LiAl$_3$N$_4$]:Eu\cite{pust2014narrow}. 
In this context, empirical rules to develop narrow-band Eu$^{2+}$-doped phosphors have been proposed.
In particular, it is believed that host crystals with high symmetry sites are favourable, leading to the development of UCr$_{4}$C$_{4}$-type Eu$^{2+}$-doped phosphors, with very narrow emission bands\cite{pust2014lial3n4,takeda2015narrow,dutzler2018alkali,liao2018learning}.

Nonetheless, the detailed physical mechanisms behind narrow Eu$^{2+}$-doped emission linewidth remains elusive.
Indeed, the empirical rules for LED phosphors design are based on ground state geometries, which are not directly linked to the emission process. 
To go beyond empirical models, one needs to consider the excited state atomic positions and focus on white LEDs working temperature range (about 450~K).\cite{xie2016nitride} 
The most important factor controlling the optical performance of LED device is the FWHM at working temperature, which is the subject of this work.

Using first-principles data on fifteen Eu$^{2+}$-doped phosphors computed in an earlier work\cite{jia2017first,jia2020erratum}, 
we calculate their FWHM using a classical and quantum mechanical model.  
We observe that quantum effects are negligible at working temperature and above. 
In this temperature range, the classical model is highly predictive and only relies on the Franck-Condon energy shifts between the ground and excited states geometries, also delivering the Stokes shift.  
Hence, for given absorption and emission energies, we search for materials with large positive difference between the Franck-Condon shifts of the excited and ground states. 
For our representative set of materials we find that the FWHM can be decreased by as much as 40~\% without changing
absorption and emission energies.

The present work is structured as follows. 
In Section~\ref{Theory}, we first describe the theoretical methods used in the analysis of emission linewidth (FWHM) of Eu$^{2+}$-doped phosphors. 
In Section~\ref{R&D}, the detailed FWHM calculation results are shown and the classical and quantum picture are compared. 
Finally, the conclusion is given in Section~\ref{con}.

\section{Theoretical approach}
\label{Theory}
The impact of lattice vibration on electronic transition has a long and rich history\cite{huang1950theory,rebane2012impurity,henderson2006optical}. 
Under the adiabatic approximation, the Schr\"odinger equation of motion can be simplified by decoupling the electron and nuclear degrees of freedom. 
After an optical electronic transition, the excited electronic density exerts a force on the nuclei.
The atoms will therefore relax to this excited-state configuration by emitting multiples phonons.
The energy difference between the the ground- and excited-state atomic configurations while the electronic state is adiabatically changed (either in the electronic ground state or excited state) is called the \textit{Franck-Condon} shift. 
One can express the total Franck-Condon shift as the sum of each vibrational eigenmodes contribution: 
\begin{equation}\label{eq:constrainE}
E_{\rm FC} =  \sum_{i=1}^{3N}E^i_{FC},
\end{equation}  
where $i$ is the mode index and $N$ the number of atoms in the unit cell. 
The usual way to handle Eq.~\eqref{eq:constrainE} is to use a single effective phonon mode of frequency $\hbar\Omega$ (for emission or for absorption).
The resulting Huang-Rhys parameter $S$ is deduced as:\cite{huang1950theory}
\begin{equation}\label{eq:Huang-Rhys}
S=  \frac{E_{\rm FC}}{\hbar\Omega}.
\end{equation} 

\begin{figure}
\begin{center}
\includegraphics[scale=0.3]{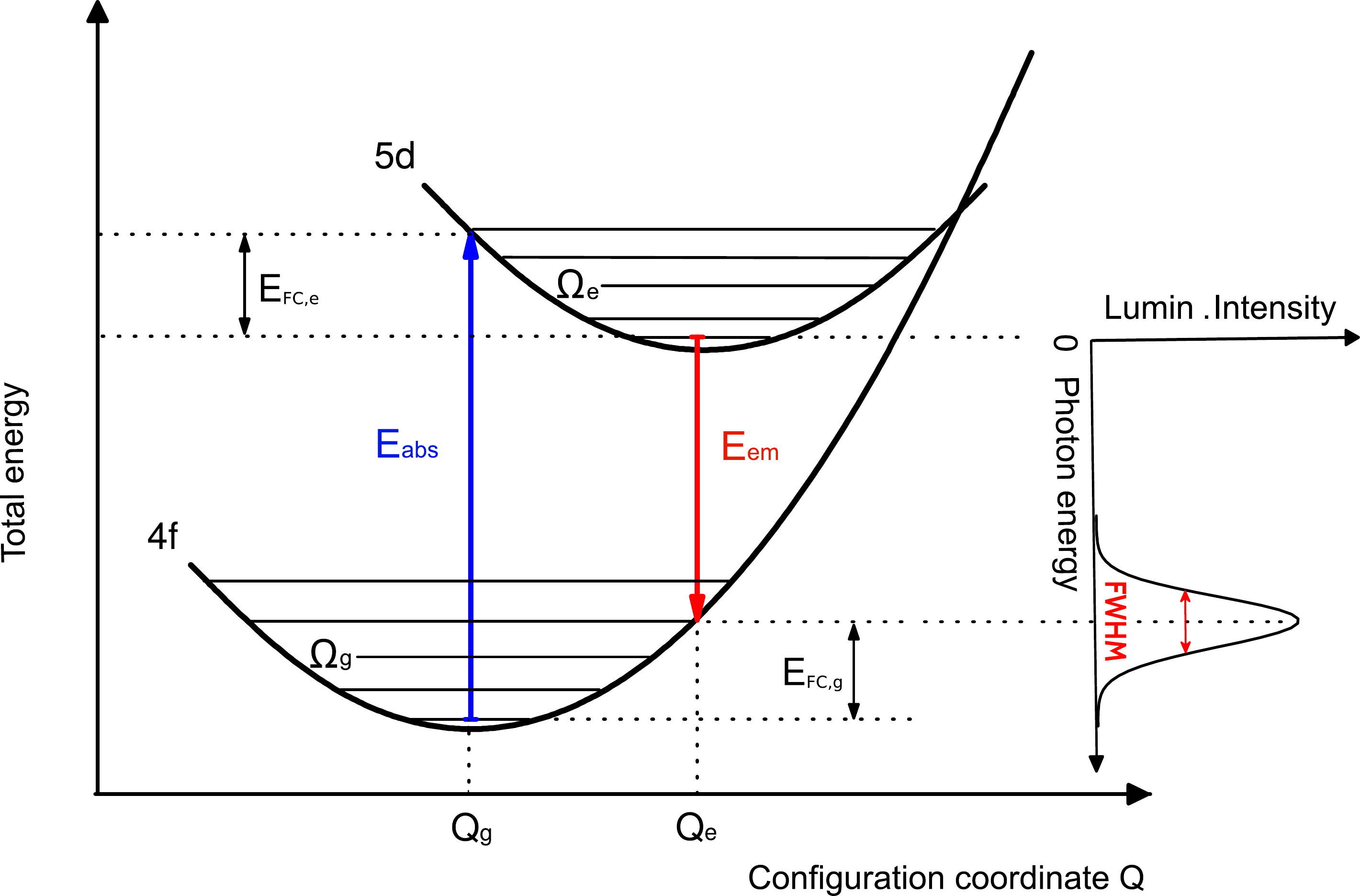}
\caption{One-dimensional Configuration Coordinate diagram for the analysis of emission full width at half maximum (FWHM) of Eu$^{2+}$-doped phosphors. 
The generalized configuration coordinate $Q$ results from the linear combination of ground state atomic positions (Q$_g$) and excited state ones (Q$_e$).
Total energies as a function of Q are reported for the electronic ground state (labeled 4f), and for the electronic excited state (labeled 5d), the absorption and emission occur between these energies at E$_{abs}$ and E$_{em}$, respectively. 
The energy lost to the lattice after light absorption defines the Franck-Condon shift E$_{FC,e}$, and similarly E$_{FC,g}$ is the energy lost to the lattice after light emission. 
$\Omega_g$ and $\Omega_e$ are the vibrational frequencies of the ground and excited state curves. 
They are linked to each spectrum shape through the so-called Huang-Rhys factor $S$, the ratio between a Franck-Condon shift and the related vibrational frequency. }\label{figure_1}
\end{center}
\end{figure}
Such approach is the backbone of the one-dimensional configuration coordinate diagram (1D-CCD) which is shown in Fig.~\ref{figure_1}.
%
It has been shown that this model works well when the electron-phonon coupling is strong (large Huang-Rhys parameter)~\cite{alkauskas2012first,alkauskas2016tutorial}.
Fortunately, the 4f-5d transition in Eu-doped phosphors belong to such category with strong electron-phonon coupling and large Franck-Condon shifts.

Within the 1D-CCD model, the Stokes shift $\Delta S$ is the sum of the Franck-Condon energy shifts in the ground and excited states: 
\begin{equation}
\Delta S = E_{\rm FC,e} + E_{\rm FC,g}.
\end{equation}

We also introduce the difference of the Franck-Condon energy shift of ground and excited states $\Delta C$:
\begin{equation}
\Delta C = E_{\rm FC,e} - E_{\rm FC,g},  
\end{equation}
which gives the relationships:
\begin{align}
E_{\rm FC,e} &= \frac{1}{2}(\Delta S + \Delta C) \label{1} \\
E_{\rm FC,g} &= \frac{1}{2}(\Delta S - \Delta C). \label{2}
\end{align}

The Franck-Condon shifts can also be expressed in terms of the Huang-Rhys parameter, following Eq.~\ref{eq:Huang-Rhys}:
\begin{align}
E_{\rm FC,g} &=  S_{\rm em} \hbar\Omega_{\rm g} \label{3} \\
E_{\rm FC,e} &=  S_{\rm abs} \hbar\Omega_{\rm e} \label{4},
\end{align}
where, for the excited state, $S_{\rm abs} = E_{\rm FC,e}/\hbar \Omega_{\rm e}$ and $\Omega_{\rm e}$ is the effective excited-state phonon frequency, 
while a similar notation applies for the ground state.

Using the 1D-CCD model, the FWHM can be calculated as~\cite{henderson2006optical,reshchikov2005,jia2017first}: 
 \begin{align}\label{eq:constrainE-T}
 W(T)  &= S_{\rm em} \hbar \Omega_{\rm g} \frac{\sqrt{8\ln2}}{\sqrt{S_{\rm abs}}} \bigg[ \coth\Big( \frac{\hbar \Omega_{\rm e}}{2 k_{\rm B}T} \Big) \bigg]^{1/2}, 
\end{align} 
where the $\coth$ term is unity at zero temperature. 

We now reformulate this equation, in order to separate
the dependence on the difference of Franck-Condon shifts from the dependence on Stokes shift and temperature. 
The emission energy only appears in conjunction with the temperature.
Inserting Eqs.~\eqref{1}-\eqref{4} in Eq.~\eqref{eq:constrainE-T}, we indeed obtain:
\begin{multline}\label{eq:constrain-C}
 W(T) = 2\sqrt{\ln2 \Delta S}\frac{(1-\Delta C/\Delta S)}{\sqrt{1+\Delta C/\Delta S}} \\
 \times \sqrt{(\hbar \Omega_{\rm e})\coth(\hbar \Omega_{\rm e}/2k_{\rm B}T)}.
\end{multline} 
 
At high temperature, $\hbar \Omega_{\rm e} << 2k_{\rm B}T$~\cite{rebane2012impurity}, Eq.~\eqref{eq:constrain-C} can be approximated by the classical expression,
where the Bose-Einstein statistics has been replaced by the Boltzmann statistics to treat the temperature effects: 
\begin{equation}\label{eq:constrain-C2}
 W_{cl}(T) = 2\sqrt{\ln2\Delta S} \frac{(1-\Delta C/\Delta S)}{\sqrt{1+\Delta C/\Delta S}}\sqrt{2k_{\rm B}T}.
\end{equation}  

The replacement of \eqref{eq:constrain-C} by Eqs.~\eqref{eq:constrain-C2} yields an error, 
that can be readily gauged by defining the ratio between these, and comparing it to one. 
We define the relative classical error $f_{rcle}$, 
a function of the ratio between temperature and effective excited-state phonon frequency, as
\begin{equation}\label{error}
 f_{rcle} \equiv 1- \sqrt{(2k_{\rm B}T/\hbar\Omega_{\rm e}) \tanh(\hbar \Omega_{\rm e}/2k_{\rm B}T)},
\end{equation} 
so that
\begin{equation}\label{eq:ratio}
 W_{cl}(T) = W(T) (1-f_{rcle}).
\end{equation} 
 
Finally, for fixed Stokes shift $\Delta S$,  the ratio between $W_{cl}(T)$ and the square root of the Stokes shift is a function of only the square root of the temperature, and the $\Delta C$/$\Delta S$
ratio,
highlighting the importance of this ratio:  
  \begin{equation}\label{eq:constrain-C3}
 W_{cl}(T)/\sqrt{\Delta S}  = 2\sqrt{\ln2} \frac{(1-\Delta C/\Delta S)}{\sqrt{1+\Delta C/\Delta S}}\sqrt{2k_{\rm B}T}.
 \end{equation}  
For a fixed temperature and Stokes shift, which are indeed determined by the working conditions and expected colour emission, the classical width can be decreased by increasing the $\Delta C$/$\Delta S$ ratio.
As the Franck-Condon shifts for both ground state and excited state are positive, the range of $\Delta C$/$\Delta S$ is between -1 and 1, which would not preclude a vanishing value for $W_{cl}(T)$ or $W(T)$.
In the next section we show the actual range of $\Delta C$/$\Delta S$ values for a set of fifteen materials.

\begin{table*}
\centering
\renewcommand\arraystretch{1.5}
\caption{The calculated FWHM and related parameters for fifteen Eu$^{2+}$-doped phosphors, some of which having two different doping sites, or ordered structures, giving eighteen different cases. Eu1 and Eu2 stand for inequivalent Eu sites in SrAl$_2$O$_4$ and Sr{[}LiAl3N4{]} compounds. M-I and M-II for CaAlSiN$_3$ stands for two ordered structures in our calculations. See the description in text and Ref.~\cite{jia2017first}. In view of further analysis, the cases are ordered with increasing $\Delta$C/$\Delta$S ratio.}
\label{data}
\begin{tabular}{lcccccccccc}\hline\hline

 Compound               & $\Delta$S & $\Delta$C & $\Delta$C/$\Delta$S  & \multicolumn{3}{c}{W(T)} & \multicolumn{2}{c}{W$_{cl}$(T)} & \multicolumn{2}{c}{W(T)/$\sqrt{\Delta S}$}  \\
        & (eV) & (eV) & --  & \multicolumn{3}{c}{(eV)} & \multicolumn{2}{c}{(eV)} & \multicolumn{2}{c}{(eV$^{1/2}$)}  \\
                          &              &               &          &     0K             & 298K   & 450K & 298K & 450K & 298K & 450K  \\ \hline
Sr{[}LiAl$_3$N$_4${]}:Eu1        & 0.133 & -0.019 & -0.143 & 0.128 & 0.179 & 0.214 & 0.17  & 0.209 & 0.466 & 0.587 \\
Sr$_2$MgSi$_2$O$_7$:Eu           & 0.585 & -0.073 & -0.126 & 0.222 & 0.357 & 0.432 & 0.347 & 0.427 & 0.454 & 0.565 \\
SrAl$_2$O$_4$:Eu2                & 0.62  & -0.071 & -0.114 & 0.198 & 0.358 & 0.435 & 0.352 & 0.432 & 0.447 & 0.552 \\
BaSi$_2$O$_2$N$_2$-Eu            & 0.558 & -0.063 & -0.112 & 0.169 & 0.336 & 0.411 & 0.333 & 0.409 & 0.446 & 0.550 \\
SrAl$_2$O$_4$:Eu1                & 0.653 & -0.071 & -0.108 & 0.233 & 0.368 & 0.446 & 0.358 & 0.44  & 0.443 & 0.552 \\
Sr$_5$(PO$_4$)$_3$Cl:Eu          & 0.313 & -0.025 & -0.078 & 0.118 & 0.239 & 0.293 & 0.237 & 0.291 & 0.424 & 0.524 \\
CaF$_2$:Eu                       & 0.212 & -0.016 & -0.077 & 0.149 & 0.205 & 0.245 & 0.195 & 0.239 & 0.424 & 0.532 \\
Ba$_3$Si$_6$O$_{12}$N$_2$:Eu     & 0.498 & -0.036 & -0.071 & 0.197 & 0.305 & 0.369 & 0.296 & 0.364 & 0.419 & 0.523 \\
CaS:Eu                           & 0.310  & -0.022 & -0.070  & 0.130  & 0.237 & 0.289 & 0.233 & 0.287 & 0.418 & 0.519 \\
CaMgSi$_2$O$_6$:Eu               & 0.522 & -0.027 & -0.052 & 0.241 & 0.315 & 0.374 & 0.295 & 0.362 & 0.408 & 0.518 \\
Sr[Mg$_3$SiN$_4$]:Eu & 0.161 & -0.008 & -0.051 & 0.086 & 0.165 & 0.202 & 0.163 & 0.200   & 0.406 & 0.503 \\
Ca{[}LiAl$_3$N$_4${]}:Eu         & 0.169 & -0.005 & -0.032 & 0.098 & 0.166 & 0.202 & 0.163 & 0.200   & 0.397 & 0.491 \\
CaAlSiN$_3$:Eu, M-I              & 0.34  & 0.002  & 0.008  & 0.098 & 0.219 & 0.268 & 0.217 & 0.267 & 0.372 & 0.460 \\
KSrPO$_4$:Eu                     & 0.612 & 0.008  & 0.013  & 0.12  & 0.291 & 0.356 & 0.289 & 0.356 & 0.369 & 0.455 \\
SrB$_4$O$_7$:Eu                  & 0.212 & 0.006  & 0.026  & 0.125 & 0.176 & 0.21  & 0.167 & 0.206 & 0.363 & 0.456 \\
CaAlSiN$_3$:Eu, M-II             & 0.31  & 0.028  & 0.088  & 0.097 & 0.186 & 0.227 & 0.184 & 0.226 & 0.33  & 0.408 \\
SrI$_2$:Eu                       & 0.212 & 0.032  & 0.154  & 0.057 & 0.138 & 0.169 & 0.137 & 0.168 & 0.298 & 0.367 \\
Sr{[}LiAl$_3$N$_4${]}:Eu2        & 0.171 & 0.052  & 0.302  & 0.067 & 0.099 & 0.12  & 0.096 & 0.118 & 0.232 & 0.290  \\   
\hline \hline 
\end{tabular}
\end{table*}

\section{Results and Discussions}
\label{R&D}

Table~\ref{data} lists the FWHM and the most relevant parameters of the fifteen Eu$^{2+}$-doped phosphors, deduced from the  $\Delta$SCF calculations of our previous work~\cite{jia2017first,jia2020erratum}. 
One has also two cases of different doping sites for the same materials, and one case of two different ordered structures for the same materials, 
increasing the total number of cases to eighteen.
Note that the accuracy of data published in Ref.~\cite{jia2017first} has been improved in Ref.~\cite{jia2020erratum}.
This is relevant for the present work, where we focus on emission linewidths, which are usually smaller than 0.4~eV,
while our previous study~\cite{jia2017first} focused mainly on the transition energy, that are observed in the region of 2-4~eV.

\begin{figure}
\begin{center}
\includegraphics[scale=0.6]{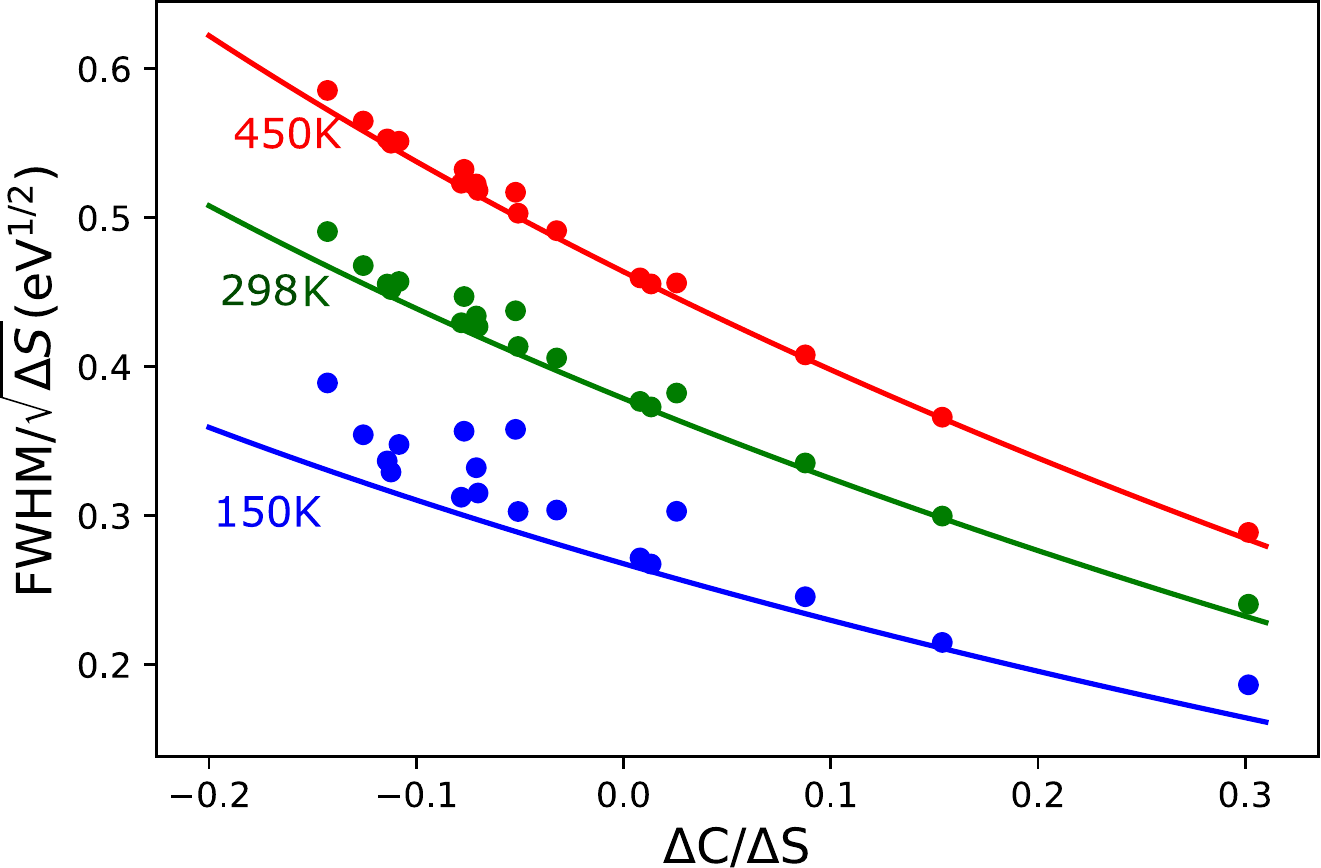}
\caption{The effect of $\Delta$C/$\Delta$S on FWHM/$\sqrt{\Delta S}$ for 150~K (blue), 298~K (green) and 450~K (red).  The lines are based on the classical approximation $W_{cl}(T)$, from Eq.~\eqref{eq:constrain-C3}, and dots are based on $W(T)$, data from Table~\ref{data} and 
Eq.~\eqref{eq:constrain-C}.}\label{figure_2}
\end{center}
\end{figure}
In Figure~\ref{figure_2} we compare the `exact' quantum mechanical (dots from Eq.~\eqref{eq:constrainE-T}) and `approximated' classical  (fitting lines from Eq.~\eqref{eq:constrain-C3}) approaches, at different temperatures (150~K, 298~K and 450~K). 
The results from the classical model provide a lower bound while the `exact' quantum model results are larger than classical.
However, as expected, the difference between the two approaches tends to vanish as the temperature increases. 
At 150K, for most case, the results are significantly different, while at the working temperature of LED devices (450~K) the difference is negligible.
We can therefore safely neglect the zero-point vibrational contribution to the FWHM calculations for our fifteen Eu$^{2+}$-doped materials. 
There are cases where the agreement between the two models is reached at lower temperature e.g. for SrI$_2$:Eu.
This is due to the very small vibrational frequencies involved.
However, in case of large vibrational frequencies, the classical model will be relevant only at higher temperature.

\begin{figure}
\begin{center}
\includegraphics[scale=0.6]{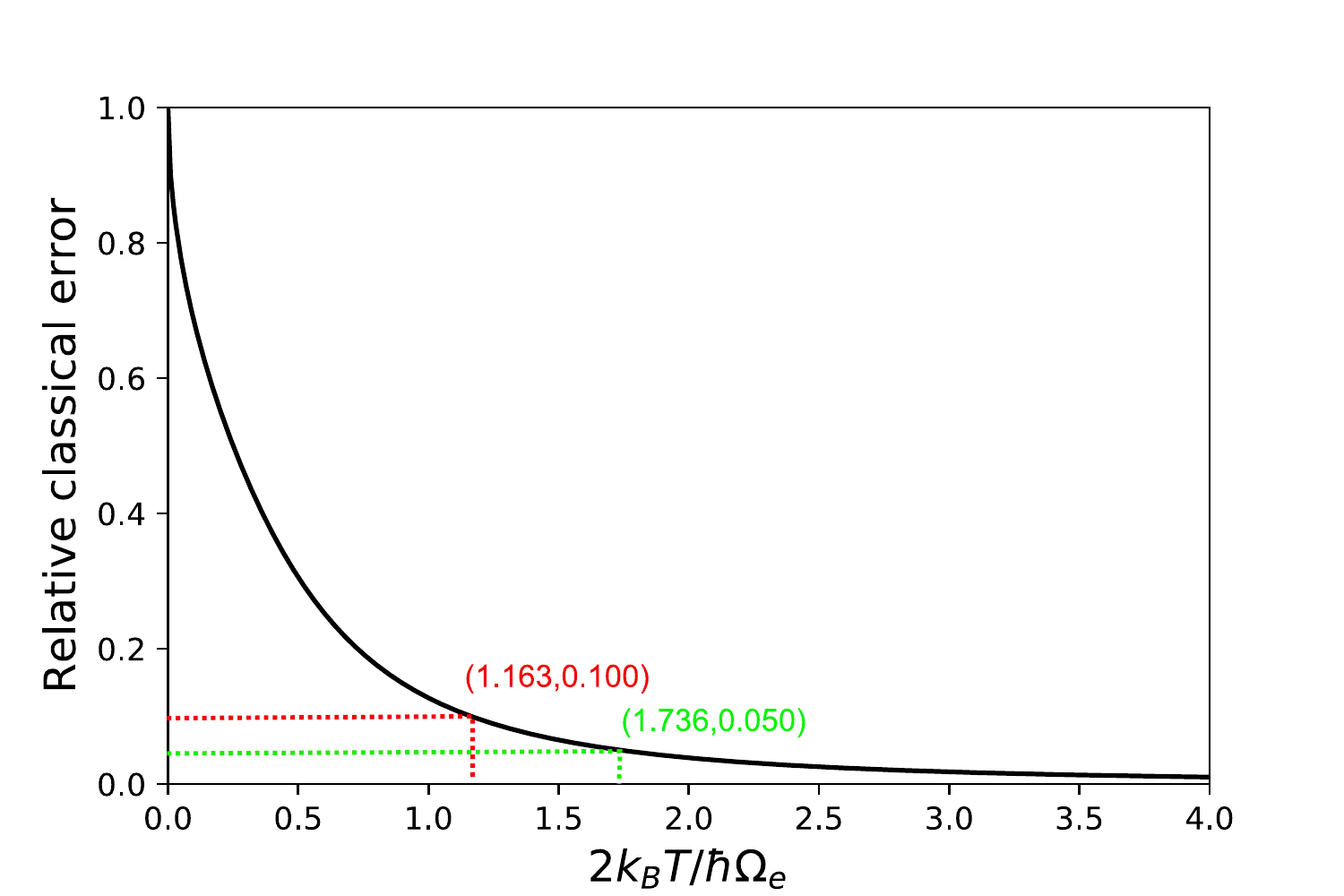}
\caption{The relative classical error $f_{rcle}$. The error is smaller than 10~\% when 2$k_B$T/$\hbar\Omega$ is higher than 1.163.} \label{figure_3}
\end{center}
\end{figure}

Therefore, we now focus on the working temperature of LED device. 
Figure~\ref{figure_3} shows the relative classical error, $f_{rcle}$, Eq.~\eqref{error}. Based on this figure, the value of 2$k_B$T/$\hbar\Omega$ can be calculated for a fixed relative classical error.
Then, the upper bound of vibrational frequency can be obtained for the working temperature temperature. 
Here, errors smaller than 5\% and 10\% are analyzed. 
Following the above procedure, such errors are obtained for 2$k_B$T/$\hbar\Omega$ equal to 1.738 and 1.163, respectively. 
The upper bound of vibrational frequency for the 450~K is finally determined to be 45~meV and 67~meV, which corresponds to 363~cm$^{-1}$ and 540~cm$^{-1}$, respectively. 
For our fifteen Eu$^{2+}$-doped phosphors such differences between the quantum and classical treatments are smaller than the upper limit of 5~\% at 450~K.
Still the chemical composition of the fifteen Eu$^{2+}$-doped phosphors are quite broad, covering the halide, oxide and nitride compounds, including the
light Li ion, and their crystal structures are quite different. 
We assume that the set of vibration frequency from the fifteen Eu$^{2+}$-doped phosphors can stand for the vibration frequency that occurs in Eu$^{2+}$-doped materials. 
Therefore, the classical  approximation should work for most LED phosphors.

We are now in a position to assess the realistic range of values for $\Delta$C/$\Delta$S. 
Indeed, in theoretical analyses of the lineshape, the hypothesis of identical curvature of the ground state and excited state total energy as a function of atomic displacements is often done, which corresponds to $\Delta$C/$\Delta$S = 0.
We have seen earlier that this ratio can mathematically vary between -1 and 1.
In Table~\ref{data}, we see that the range for $\Delta$C/$\Delta$S for the eighteen different cases is between -0.143 and +0.302, resp., with extremal values for Sr[LiAl$_3$N$_4$]:Eu1 and Sr[LiAl$_3$N$_4$]:Eu2, resp. 
This corresponds, resp., to an increase of the $W_{cl}(T)/\sqrt{\Delta S}$ ratio by 23\%, or, more interestingly, a decrease of this ratio by nearly 40\% with respect to the $\Delta$C/$\Delta$S = 0 value.
For twelve cases in our set, the $\Delta$C/$\Delta$S ratio is negative, which induces widening of the FWHM. 
$\Delta$C/$\Delta$S values lower than -0.1 are even present for five materials, with increase of  $W_{cl}(T)/\sqrt{\Delta S}$ by more than 15\%.
For the remaining six cases, the FWHM is decreased, although this decrease is usually quite small, since only for two cases (SrI$_2$:Eu and the above-mentioned Sr[LiAl$_3$N$_4$]:Eu2), $\Delta$C/$\Delta$S is higher than 0.1 with a decrease larger than 15\%.

We can now consider the narrow LED red emitter target
from the United States Department of Energy. 
This target FWHM of 30nm, that should be obtained for red emission,
corresponds to about 0.1 eV FWHM energy. 
If we take 0.3 as maximum value of $\Delta$C/$\Delta$S,
as indicated from our survey of fifteen Eu$^{2+}$-activated phosphors, and if we consider the working temperature of 450K, the $\Delta$S for FWHM = 0.1eV is about 0.125eV, or 1000~cm$^{-1}$.  
This value provides an upper bound for the $\Delta$S of Eu$^{2+}$ emission phosphors, that can meet the requirement of narrow LED converters from the United States Department of Energy. 
The $\Delta$S and FWHM are of similar magnitude,
which indicates that the absorption and emission
spectra will overlap significantly, with undesirable side effects. 
A smaller value of $\Delta$C/$\Delta$S corresponds also to a smaller needed $\Delta$S, with larger overlap between 
absorption and emission spectra.

Of course, this conclusion might be questioned, as being an outcome of the simple 1D-CCM. 
For more elaborate calculation of FWHM, taking into account the variety of vibrational mode frequencies, the picture might change somehow.
Still, to find a narrow Eu$^{2+}$ emission phosphor, the ratio of $\Delta$C/$\Delta$S can be a very useful descriptor.
The larger this ratio, for fixed Stokes shift and temperature,
the smaller the FWHM.
This might be used as a design rule in the high-throughput 
computational search for new phosphors.

\begin{figure}
\begin{center}
\includegraphics[scale=0.3]{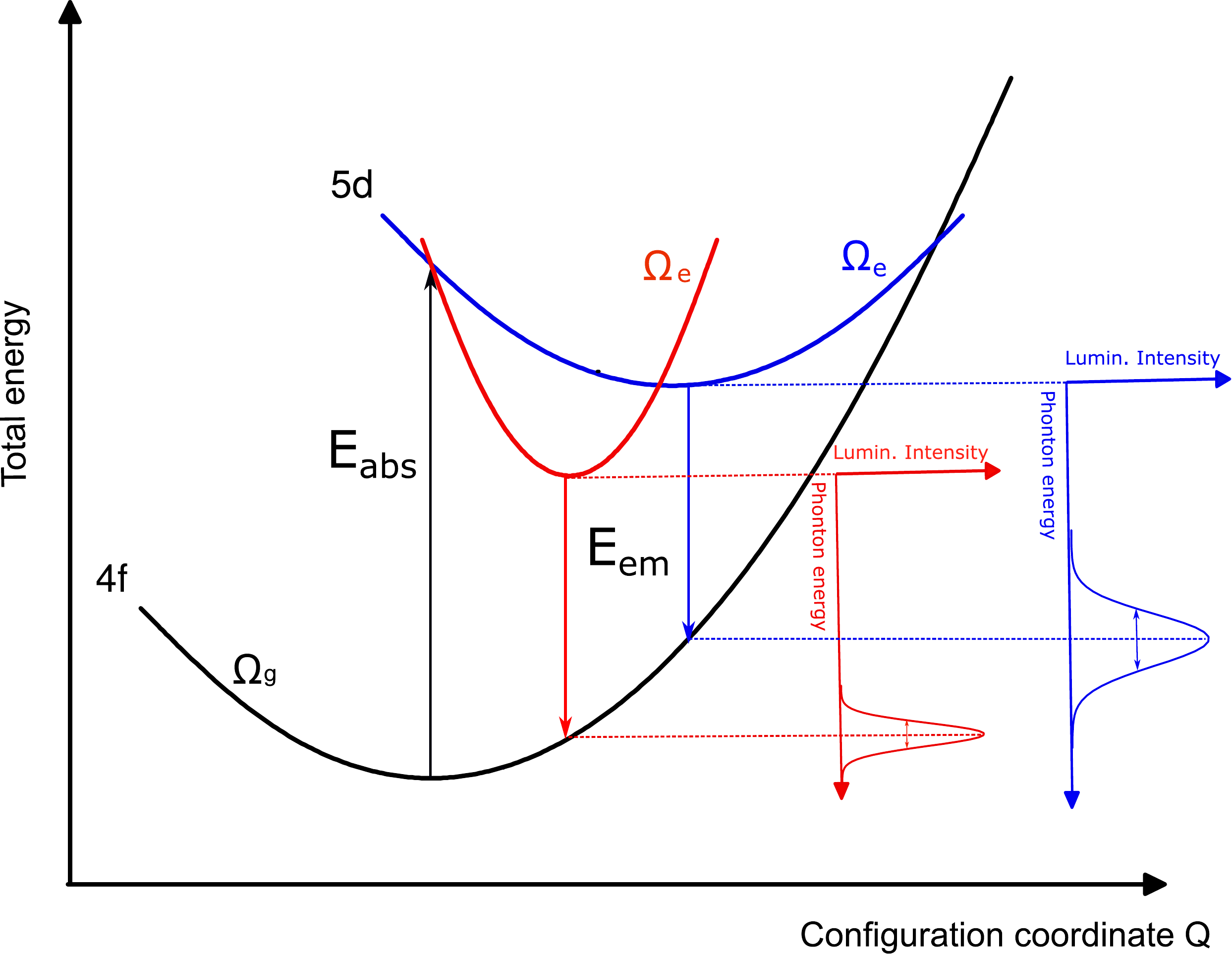}
\caption{Representation of the effect of $\Delta$C on FWHM, under the condition of fixed $\Delta$S and emission frequency. The blue curves correspond to vanishing $\Delta$C, while the red curves
correspond to large positive $\Delta$C.}\label{figure_4}
\end{center}
\end{figure}

\section{Conclusion}
\label{con}

In summary, we have studied the full width at half maximum of the emission spectrum of fifteen Eu$^{2+}$-doped LED phosphors, using the one-dimensional configuration coordinate model, from formulas based on quantum and classical statistics. 
%
At the LED working temperature, 450~K, the FWHM of Eu$^{2+}$-doped LED phosphors can be described classically,  quantum effect of lattice vibrations can be safely neglected. 

At fixed emission energy and Stokes shift, the difference of Franck-Condon energy shifts is the key factor in the determination of the FWHM. 
A positive Franck-Condon energy shift between excited and ground state will induce a narrower emission band than a negative one.
Taking as reference the case where the Franck-Condon shifts of the ground and excited states are equal,
in our representative set of materials, the FWHM can be decreased by as much as 40~\%, if the Franck-Condon shift of the excited state is larger than the one of the ground state. 
By contrast, the FWHM might be increased by 15~\% in the opposite case. 
The high-throughput 
computational search for new phosphors
might rely on the ratio between the 
difference of Franck-Condon shift of the excited state and the one of the ground state, divided by the Stokes shift.

\vspace{0.2cm}

\begin{flushleft}
\begin{large}
\textbf{Acknowledgement}
\end{large}
\end{flushleft}

\vspace{0.1cm}

This work, done in the framework of ETSF (project number 551), has been supported by the FRS-FNRS Belgium through a Charg\'e de recherches fellowship (Y. Jia) and the PdR Grant No. T.0103.19 - ALPS (X. Gonze). S.P. acknowledge support from the European Unions Horizon 2020
Research and Innovation Programme, under Marie Sk\l{}odowska-Curie Grant Agreement SELPH2D No.~839217. Computational resources have been provided by supercomputing facilities of the Universit\'e catholique de Louvain (CISM/UCL) and the Consortium des Equipements de Calcul Intensif en F\'ed\'eration Wallonie Bruxelles (CECI) funded by the Fonds de la Recherche Scientifique (FRS-FNRS Belgium) under Grant No. 2.5020.11.
The present research benefited from computational resources made available on the Tier-1 supercomputer of the 
F\'ed\'eration Wallonie-Bruxelles, infrastructure funded by the Walloon Region under the grant agreement No. 1117545.

\bibliography{Eu_FWHM}

\end{document}